\documentclass{iau} 
\usepackage{graphicx}

\title[Star Formation History of IC\,1613 Dwarf Galaxy] 
{Recovering the Star Formation History of IC\,1613 Dwarf Galaxy Using Evolved Stars}

\author[Seyed Azim Hashemi, Atefeh Javadi \& Jacco Th. van Loon]  
{Seyed Azim Hashemi$^{1,2}$, Atefeh Javadi$^2$
 \and Jacco Th. van Loon$^3$}

\affiliation{$^1$Department
of Physics, Sharif University of Technology, Tehran, 11155-9161, Iran\\ email: {\tt hashemi\_seyedazim@physics.sharif.edu} \\[\affilskip]
$^2$School of Astronomy, Institute for Research in Fundamental Sciences (IPM), Tehran, 19395-5531, Iran \\email: {\tt atefeh@ipm.ir}

$^3$Lennard-Jones Laboratories, Keele University, ST5 5BG, UK \\email: {\tt j.t.van.loon@keele.ac.uk }}

\pubyear{2018}
\volume{344}  
\jname{Dwarf Galaxies: From the Deep Universe to the Present}
\editors{A.C. Editor, B.D. Editor \& C.E. Editor, eds.}
\begin{document}

\maketitle

\begin{abstract}
Determining the star formation history (SFH) is key to understand
the formation and evolution of dwarf galaxies. Recovering the SFH in resolved
galaxies is mostly based on deep colour--magnitude diagrams (CMDs),
which trace the signatures of multiple evolutionary stages of their
stellar populations. In distant and unresolved galaxies, the integrated
light of the galaxy can be decomposed, albeit made difficult by an age--metallicity
degeneracy. Another solution to determine the SFH of resolved galaxies is
based on evolved stars; these luminous stars are the most accessible tracers
of the underlying stellar populations and can trace the entire SFH.
Here we present a novel method based on long period variable (LPV) evolved
asymptotic giant branch (AGB) stars and red supergiants (RSGs). We applied
this method to reconstruct the SFH for IC\,1613, an irregular dwarf galaxy
at a distance of 750 kpc. Our results provide an independent confirmation
that no major episode of star formation occurred in IC\,1613 over the past 5 Gyr.

\end{abstract}

\firstsection
\section{Introduction}
Because of observational constraints, different methods have been
developed to find the star formation history (SFH) of galaxies.
Recovering the SFH in  resolved galaxies is mostly based on color--magnitude
diagrams (CMDs), where we can trace the signatures of  most of the stellar
population using  high resolution  observations
(e.g., Tosi, Greggio \& Focardi 1989; Gallart et al.\ 1996; Dolphin 1997, 2002).
As a matter of fact, because of the limited  spatial resolution, this kind of
analysis will limit us  to a few dozen galaxies, mostly in our group (within $\sim$2
Mpc, Ruiz-Lara et al.\ 2015).  In the distant and unresolved galaxies, analysis
based on the individual stars is unfeasible, except possibly for the very
brightest. Therefore, we are limited to the integrated light of galaxies
(Ganda et al.\ 2009;  P\'erez \& S\'anchez-Bl\'azquez 2011). The integrated
light of the galaxy can be decomposed, albeit made difficult by an age--metallicity
degeneracy (e.g., Pickles 1985; Bica 1988; MacArthur et al. 2004). This method is
relatively successful in deriving the SFHs of simple systems, but still have some
limitations to give the detailed SFHs of galaxies with complex stellar 
compositions (Ruiz-Lara et al.\ 2015).

Another solution to determine the SFH of resolved galaxies is based on evolved
stars (Javadi et al. 2011a,c, 2013, 2015a,b)
these luminous stars are the most accessible tracers of the underlying stellar populations and can
trace the entire SFH. In this paper we are going to apply a method recently developed in Javadi,
van Loon \& Mirtorabi (2011b), in which we  use  long period variables (LPVs) to find the SFH.
LPVs  are mostly evolved asymptotic giant branch (AGB) stars  (e.g., Fraser et al.\ 2005, 2008;
Soszy\'nski et al.\ 2009; Boyer et al.\ 2017; Goldman et al.\ 2019; Yuan et al.\ 2018),
 and in addition to their luminosity, their variability also helps us
to identify them.  This method already has been used successfully to derive SFHs of some nearby
galaxies (Javadi et al.\ 2011b, 2016, 2017; Rezaeikh et al.\ 2014; Hamedani Golshan et al.\ 2017;
 Hashemi et al.\ 2017; Hashemi et al.\ 2019).
In the following, we will apply
this technique to find the SFH of IC\,1613 dwarf galaxy.

\section{Data and Method}

IC\,1613 is an isolated,  irregular dwarf galaxy within the Local Group
at distance of 750 kpc (Menzies et al.\ 2015). Its vicinity, inclination angle
(Lake \& Skillman 1989), and low foreground reddening
(Schlegel, Finkbeiner \& Davis 1998), makes it a great  opportunity
for the study of its stellar populations. Our sample to find
the SFH of IC\,1613 contains 53 evolved stars in the field of view
of about 200 arcmin square in the central part of the galaxy. These
53 stars have been obtained using a combination of data, that were
published by Menzies et al.\ (2015), Boyer et al.\ (2015).
Previous application of our method (Javadi et al.
2011b, 2017; Rezaeikh et al.\ 2014; Hamedani Golshan et al.
2017) was based on confirmed LPVs.  In the case at hand, though,
the limited cadency of the DUSTiNGS survey (Boyer et al. 2015) and
the limited depth of the Menzies et al.\ (2015) survey will have
led to LPVs being missed. 
Therefore, we also add AGB and red supergiant (RSG) candidate stars that 
are expected to be LPVs to our confirmed LPVs listed in Menzies et al.\ (2015).
These candidates are:

1-Extreme-AGB (x--AGB) stars (from Boyer et al.\ 2015) and large amplitude variable
(LAV) stars (Menzies et al.\ 2015) that do not have determined period but
are expected to be LPVs in the
end of AGB phase (e.g., Schr\"oder et al.\ 1999; Fraser
et al.\ 2008; Soszy\'nski et al.\ 2009).

2- RSG stars (Menzies et al.\ 2015) without determined period,
but must be good candidates for being LPVs.

Using Padova evolutionary models and assuming constant metallicities,
we can find mass, age and pulsation duration (duration  that  stars are
in LPV phase) of LPVs, and then we sort the LPVs according to their ages
and divide them into several bins. For different bins, with  specified
intervals in age and mass, we can find the star formation rate (SFR) as follows:

\begin{equation}
\xi(t) = \frac{\int_{\rm min}^{\rm max}f_{\rm IMF}(m)m\ dm}
{\int_{m(t)}^{m(t+dt)}f_{\rm IMF}(m)\ dm}\ \frac{dn^\prime(t)}{\delta t},
\label{eq:eq1}
\end{equation}

where the m is  mass, $f_{\rm IMF}(m)$ is  Kroupa initial mass function (IMF) (Kroupa 2001),
$dn^\prime$ is the observed LPVs in each bins, and $\delta t$ is pulsation duration.

The statistical error bar for each bin can be derived from Poisson distribution:

\begin{equation}
\sigma_{\xi(t)}=\frac{\sqrt{N}}{N}\xi(t),  
\label{eq:eq2}
\end{equation}

where N is the number of stars in each age bin. 

\section{Results and Discussion}

\begin{figure}[t]
\begin{center}
 \includegraphics[width=3.6in]{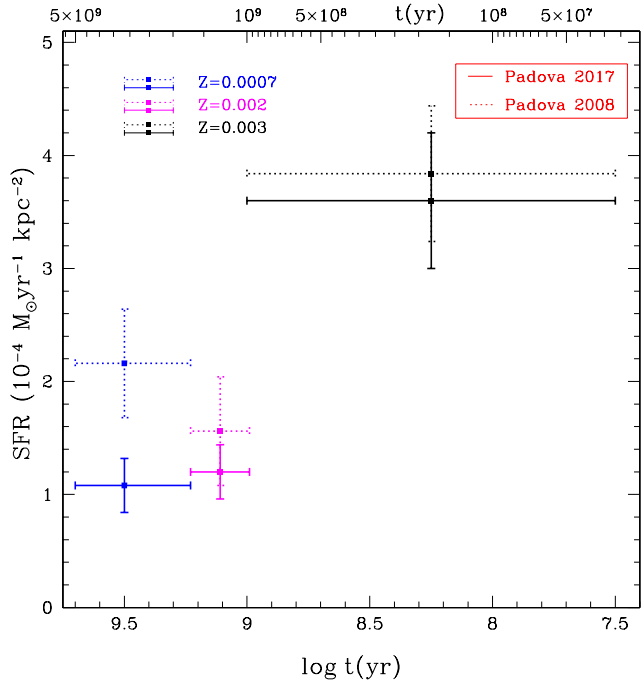} 
 \caption{The SFH in IC\,1613 with the adoption of age--metallicity relation 
 that derived by Skillman et al.\ (2014), using the pulsation
durations from Marigo et al.\ (2017) (solid) and Marigo et al.\
(2008) (dotted).}
   \label{fig1}
\end{center}
\end{figure}

Using this new technique, we estimate SFRs in IC\,1613 over the broad
time interval from 30 Myr to $\sim5$ Gyr ago. Because of the
metallicity variations over time as a result of the chemical
evolution due nucleosynthesis and mass loss, we adopt the linear age--metallicity
relation (AMR) from Skillman et al.\ (2014) for values
$4\times10^{-4}<Z<4\times10^{-3}$ (corresponding to 13 Gyr ago $<t<$ now).
Therefore, we applied $Z=0.003$ for the last Gyr
($\log t({\rm yr})<9$), $Z=0.002$ for 1 Gyr $<t< 2$ Gyr ($9<\log t({\rm
yr})<9.3$) and $Z=0.0007$ for $t>2$ Gyr ($\log t({\rm yr})>9.3$). 
Considering these metallicities, our result is shown in Fig.\,\ref{fig1}. We obtained
a mean value of the SFR across IC\,1613 over the last Gyr of
$\xi=(3.0\pm0.5)\times10^{-4}$ M$_\odot$ yr$^{-1}$ kpc$^{-2}$
, in excellent agreement with Skillman et al.\ (2014), who found
$\xi\sim3.4\times10^{-4}$ M$_\odot$ yr$^{-1}$ kpc$^{-2}$.

Our results, in the older time intervals ($9<\log t({\rm yr})<9.6$), are between 30\%
to 50\% of the values  derived by Skillman et al.\ (2014). We think that these
lower rates are mainly because of incompleteness in our data (Menzies et al.\ 2015),
however, estimation of pulsation duration ($\delta t$)  also can be a possible
reason for that. In  Fig.\,\ref{fig1}, solid lines and symbols show the results
which are obtained using 2017 Padova models  (based on Marigo et al.\ 2017)
and dotted lines and symbols show the resulting SFRs when we use the pulsation
duration estimated using 2008 Padova models (based on Marigo et al.\ 2008).
In fact, using pulsation duration from 2008 Padova models  leads to  higher
star formation rates, which are more compatible to rates that are presented
by Skillman et al.\ (2014).

\section{Conclusion}

In this paper using only 53 evolved stars, we applied a novel method to find the SFH of IC\,1613 dwarf galaxy.
Our results show that the average  star formation rate of the galaxy during the last Gyr 
is $\sim3\times10^{-4}$ M$_\odot$ yr$^{-1}$ kpc$^{-2}$. Furthermore, the uniform SFH without any dominant epoch 
of star formation over the past 5 Gyr suggests that IC\,1613 has evolved in an isolated environment at least 
in the past 5 Gyr.

\end{document}